\documentclass[11 pt,a4paper]{article}
\usepackage[utf8]{inputenc}
\usepackage{amsmath}
\usepackage{amsfonts}
\usepackage{amssymb}
\usepackage{geometry}
\usepackage{pgfplots}
\usepackage{tikz}
\usepackage{subcaption}
\usepackage{amsmath}
\usepgfplotslibrary{fillbetween}
\usetikzlibrary{patterns}
\usepackage{indentfirst}
\usepackage{amsmath}
\usepackage{booktabs}
\usepackage{pgfplotstable}
\usepackage[english]{babel}
\usepackage{amsthm}
\newtheorem{theorem}{Theorem}

\author{Dr. Jacques Balayla MD, MPH, CIP, FRCSC\footnote{To whom correspondence should be addressed: Dr. Jacques Balayla MD, MPH, CIP, FRCSC. Quilligan Scholar. e-mail: jacques.balayla@mail.mcgill.ca. Department of Obstetrics and Gynaecology. Clinique Medic Elle, Montreal, Quebec, Canada}}

\title{\LARGE Derivation of Generalized Equations for the Predictive Value of Sequential Screening Tests}

\date{}

\begin{document}
\newgeometry{top=2.5cm, bottom=2.5cm}
\maketitle  
\begin{abstract}

Using Bayes' Theorem, we derive generalized equations to determine the positive and negative predictive value of screening tests undertaken sequentially. Where a is the sensitivity, b is the specificity, $\phi$ is the pre-test probability, the combined positive predictive value, $\rho(\phi)$, of $n$ serial positive tests, is described by: 
\begin{large}
\begin{equation}
\rho(\phi) = \frac{\phi\displaystyle\prod_{i=1}^{n}a_n}{\phi\displaystyle\prod_{i=1}^{n}a_n+(1-\phi)\displaystyle\prod_{i=1}^{n}(1-b_n)} 
\end{equation}
\end{large}
If the positive serial iteration is interrupted at term position $n_i-k$ by a conflicting negative result, then the resulting negative predictive value is given by:
\begin{large}
\begin{equation}
\psi(\phi) = \frac{[(1-\phi)b_{n-}]\displaystyle\prod_{i=b_{1+}}^{b_{(n-1)+}}(1-b_{n+})}{[\phi(1-a_{n-})]\displaystyle\prod_{i=a_{1+}}^{a_{(n-1)+}}a_{n+}+[(1-\phi)b_{n-}]\displaystyle\prod_{i=b_{1+}}^{b_{(n-1)+}}(1-b_{n+})}
\end{equation}
\end{large}
Finally, if the negative serial iteration is interrupted at term position $n_i-k$ by a conflicting positive result, then the resulting positive predictive value is given by:
\begin{large}
\begin{equation}
\lambda(\phi)= \frac{\phi a_{n+}\displaystyle\prod_{i=a_{1-}}^{a_{(n-1)-}}(1-a_{n-})}{\phi a_{n+}\displaystyle\prod_{i=a_{1-}}^{a_{(n-1)-}}(1-a_{n-})+[(1-\phi)(1-b_{n+})]\displaystyle\prod_{i=b_{1-}}^{b_{(n-1)-}}b_{n-}}
\end{equation}
\end{large}
\\
\
The aforementioned equations provide a measure of the predictive value in different possible scenarios in which serial testing is undertaken. Their clinical utility is best observed in conditions with low pre-test probability where single tests are insufficient to achieve clinically significant predictive values and likewise, in clinical scenarios with a high pre-test probability where confirmation of disease status is critical.
\end{abstract} 
\restoregeometry
\newpage

\section{Predictive Value of Screening Tests}
The concepts of positive $\rho(\phi)$ and negative $\sigma(\phi)$ predictive value lie at the core of clinical diagnostics \cite{balayla2020prevalence}. The positive predictive value is defined as the probability that an individual with a positive screening test does indeed have the disease tested for. Conversely, the negative  predictive value is the probability that an individual with a negative screening test does not in fact have the disease tested for \cite{marshall1989predictive}. In technical terms, where T is true, F is false, P is positive, N is negative, $a$ is the sensitivity, $b$ is the specificity and $\phi$ is the disease prevalence/pre-test probability, the positive predictive value $\rho(\phi)$ of a screening test is defined following Bayes' Theorem as:

\begin{large}
\begin{equation}
PPV = \rho(\phi) = \frac{TP}{TP + FP} = \frac{a\phi}{a\phi+(1-b)(1-\phi)}
\end{equation}
\end{large}

Likewise, the negative predictive value $\sigma(\phi)$ is defined as:
\

\begin{large}
\begin{equation}
NPV = \sigma(\phi) = \frac{TN}{TN + FN} = \frac{b(1-\phi)}{\phi(1-a)+b(1-\phi)}
\end{equation}
\end{large}

Moreover, we denote the false negative rate (FNR) as the compliment of the sensitivity, and the false positive rate (FPR) as the compliment of the specificity \cite{pewsner2004ruling}, such that:
\begin{large}
\begin{equation}
FNR = 1-a = \frac{FN}{FN + TP} 
\end{equation}
\end{large}
\begin{large}
\begin{equation}
FPR = 1-b = \frac{FP}{FP + TN}
\end{equation}
\end{large}
\\
\
For the purposes of the concepts henceforth discussed, it is critical to understand that the FNR represents actual cases misdiagnosed as being disease-free, and the FPR represents disease-free cases diagnosed as having the condition in question. Since it is the actual underlying status of the patient that is relevant to the predictive value equations then we need to consider true negatives and false positives (and vice-versa) as belonging to the same group since the underlying disease status is the same in each case. These have the following epidemiological definitions:

\begin{align*}
True \ negatives =
\begin{cases} 
      TN = specificity = b\\
       FP = 1-specificity = 1-b\\
       \end{cases}
       \\ 
True \ positives =
\begin{cases} 
      TP = sensitivity = a\\
       FN = 1-sensitivity = 1-a\\  
\end{cases}
\end{align*}
\\
\

\subsection{Properties of Predictive Curves}

As per Bayes' theorem, $\rho(\phi)$ and $\sigma(\phi)$ are prevalence-dependent. It thus follows that the reliability of screening tests is prevalence-dependent as well. We have shown in previous work that a testing system's predictive ability can tolerate significant drops in prevalence/pre-test probability, up until a certain well-defined inflection point known as the \textit{prevalence threshold}, $\phi_e$, below which the reliability of a positive screening test drops precipitously \cite{balayla2020bayesian}. This is independent of the adequacy and reliability of the test and thus implies inherent Bayesian limitations to the screening process itself.
\begin{large}
\begin{equation}
\phi_e=\frac{\sqrt{a\left(-b+1\right)}+b-1}{(a+b-1)}
\end{equation}
\end{large}
Fortunately, sequential testing overcomes the aforementioned Bayesian limitations \cite{balayla2020bayesian}. Indeed, in order to attain a desired positive predictive value of $\rho$ that approaches $k$, the number of serial positive test iterations needed, for a $single$ $test$ carried out repeatedly, as per the $tablecloth$ \cite{balayla2020bayesian} function is:
\begin{large}
\begin{equation}
n_i =\lim_{\rho \to k}\left\lceil\frac{ln\left[\frac{\rho(\phi-1)}{\phi(\rho-1)}\right]}{ln\left[\frac{a}{1-b}\right]}\right\rceil 
\end{equation}
\end{large}
where $n_i$ = number of testing iterations necessary to achieve $\rho$, the desired positive predictive value,  and $k$ = constant such that 0 $<$ $k$ $<$ 1.

\section{Orthogonal Testing}

We define orthogonal testing in the screening context as the serial application of statistically independent screening tests to enhance the positive and/or negative predictive value. In comparison to the tablecloth function, which applies a singular test serially, orthogonal testing is carried out when two or more different tests that screen for the same condition are carried out in subsequent order. Using Bayes' Theorem we can derive a formula to determine the positive and negative predictive values using orthogonal testing. Where D = disease, $\neg$D = no disease, T = positive test, $\neg$T = negative test, the expression of Bayes' theorem in generalized terms for the presence of disease in the presence of a positive test is the following:
\begin{large}
\begin{equation}
P(D|T) = \frac{P(T|D) P(D)}{ P(T|D) P(D) + P(T|\neg D) P(\neg D)}
\end{equation}
\end{large}
Likewise, for the absence of disease with a negative test, we obtain:
\begin{large}
\begin{equation}
P(\neg D|\neg T) = \frac{P(\neg T|\neg D) P(\neg D)}{ P(\neg T|\neg D) P(\neg D) + P(\neg T|D) P(D)}
\end{equation}
\end{large}
\begin{theorem}
Let $T_1,T_2,...,T_n$ denote n $independent$, $positive$ tests, such that \\
$P(T_1,T_2,...T_n) = [P(T_1|D)P(T_2|D)...P(T_n|D)]P(D) +  [P(T_1|\neg D)...P(T_n|\neg D)]P(\neg D)$. Since the sensitivity a = $P(T|D)$ and specificity b = $P(\neg T| \neg D)$ such that \\ $P(T|\neg D)$ = 1-b, then, as per Bayes' Theorem, the expression for $P(D|T_1,T_2,...,T_n$) is:

\begin{align*}
P(D|T_1,T_2,...T_n) &= \dfrac{P(T_1,T_2...T_n|D)P(D)}{P(T_1,T_2...T_n|D)P(D) + P( T_1,T_2...T_n|\neg D)P(\neg D)}
\end{align*}
\end{theorem}
\

In terms of screening parameters, the above equation therefore becomes:
\begin{equation}
\rho(\phi)= \frac{a_1...a_n\phi}{ a_1...a_n\phi+(1-b_1)...(1-b_n )(1-\phi)} = \frac{\phi\displaystyle\prod_{i=1}^{n}a_n}{\phi\displaystyle\prod_{i=1}^{n}a_n+(1-\phi)\displaystyle\prod_{i=1}^{n}(1-b_n)}
\end{equation}
\\
\
where $a_n$ is the test-specific sensitivity, $b_n$ is the test-specific specificity and $n$ is the number of serial tests undertaken. The above follows from the law of total probability. Of importance, this equation allows for the calculation of the positive predictive value only in cases of serial positive tests carried out orthogonally. Visually, we can observe the effects that orthogonal testing with serial positive tests has on the positive predictive value curve for three independent, positive tests $T_1$, $T_2$, $T_3$ of varied sensitivity and specificity parameters:

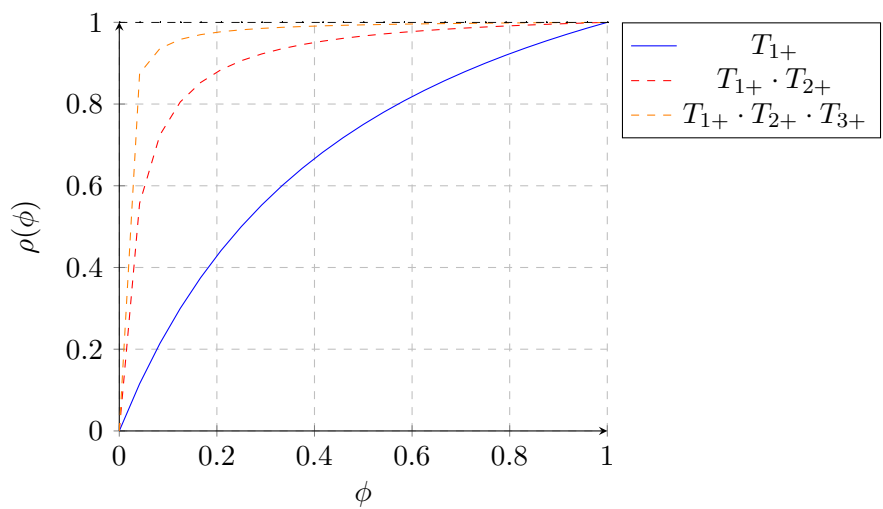
\begin{center}
\begin{tikzpicture}
	\begin{axis}[
    axis lines = left,
    xlabel = $\phi$,
	ylabel = {$\rho(\phi)$},    
     ymin=0, ymax=1,
    legend pos = outer north east,
     ymajorgrids=true,
     xmajorgrids=true,
    grid style=dashed,
    width=8cm,
    height=7cm,
     ]
	\addplot [
	domain= 0:1,
	color= blue,
	]
	{(0.60*x)/((0.60*x+(1-0.80)*(1-x))};

\addplot [	
    dashed,
	domain= 0:1,
	color= red,
	]
	{0.60*(0.81*x)/(0.60*(0.81*x+(1-0.86)*(1-0.80)*(1-x))};
	
	\addplot [
	dashed,	
	domain= 0:1,
	color= orange,
	]
	{0.70*0.60*(0.81*x)/(0.70*0.60*(0.81*x+(1-0.86)*(1-0.80)*(1-0.82)*(1-x))};
\addplot+ [
dashed,
domain= 0:1,	
color = black,
mark size = 0pt
 ]
	{1};
	\addlegendentry{$T_{1+}$}
	\addlegendentry{$T_{1+} \cdot T_{2+}$}
	\addlegendentry{$T_{1+} \cdot T_{2+} \cdot T_{3+}$}
	\end{axis}
\end{tikzpicture}
\end{center}
From the above relationship, we observe that as serial different positive tests are accounted for, we reach the limit of $\rho(\phi)$ = 1, as set by the fundamental theorem of screening \cite{balayla2020prevalence}: 
\begin{equation}
\lim_{n \to \infty}\rho_n(\phi) = \lim_{a+b \to 2}\int_{0}^{1}{\rho(\phi)d\phi} = 1
\end{equation}
\newpage
\section{Sequential Testing}
When faced with a given clinical scenario amenable to screening, one may calculate $n_i$ to determine the number of serial positive tests needed to obtain a desired positive predictive value, particularly if the prevalence or pre-test probability is low. As discussed, the latter can be accomplished by utilizing a single test serially, in which case the $tablecloth$ equation (9) applies, or by undertaking orthogonal testing with two or more tests serially, in which case the generalized equation in (12) applies. The level of predictive value opted for will depend on the particular disease being screened for, and the degrees of comfort with Type 1 and Type 2 errors that one is willing to incur \cite{mayo2011error}. A type I error is the rejection of a true null hypothesis (also known as a false positive finding or conclusion), while a type II error is the non-rejection of a false null hypothesis (also known as a false negative finding or conclusion)\cite{mayo2011error}. What happens if during a finite sequence S of positive test ($t$) iterations, be it from a single test or different tests, one stumbles upon a negative test prior to attaining the desired positive predictive value?

\begin{center}
S = $\{t_\pm\}_{t=0}^{n_i} = \lbrace t_{1+}, t_{2+}, t_{3-},...n_i \rbrace$
\end{center}

First, from a purely probabilistic standpoint, in cases where $\sigma(\phi)$ $>>$ $\rho(\phi)$, it follows that in the absence of strictly positive serial results, we consider the presence of a negative test as dominant and come to the conclusion that the individual tested is in fact unlikely to have the disease. That said, what if $\rho(\phi)$ $\backsim$ $\sigma(\phi)$ after $n_i$ - k tests? With contradictory orthogonal tests we can calculate the negative predictive value through the following formula, $\psi(\phi)$, derived from Bayes' Theorem by modifying the equation in (5) in cases where $\sigma(\phi)$ $>$ $\rho(\phi)$. The derivation stems directly from Theorem 1 as above.
\\
\
\begin{large}
\begin{equation}
\psi(\phi) = \frac{(1-b_{1+})(1-b_{2+})...(1-b_{n+})(1-\phi)b_-}{a_{1+}a_{2+}...\phi(1-a_-)+(1-b_{1+})(1-b_{2+})...(1-b_{n+})(1-\phi)b_{-}}
\end{equation}
\end{large}

where $a_{x+}$ and $b_{x+}$ are the sensitivity and specificity of the $xth$ positive test and $a_{x-}$ and $b_{x-}$ are the like parameters of the interjecting negative test. Since in this case we undertake serial testing to enhance the positive predictive value, it is implied that $\sigma(\phi)$ $>$ $\rho(\phi)$. Therefore, the above equation $\psi(\phi)$ is a modification of $\sigma(\phi)$ and provides a revised measure of the negative predictive value. The above equation is simplified as:
\begin{equation}
\psi(\phi) = \frac{[(1-\phi)b_{n-}]\displaystyle\prod_{i=b_{1+}}^{b_{n-1+}}(1-b_{n+})}{[\phi(1-a_{n-})]\displaystyle\prod_{i=a_{1+}}^{a_{n-1+}}a_{n+}+[(1-\phi)b_{n-}]\displaystyle\prod_{i=b_{1+}}^{b_{n-1+}}(1-b_{n+})}
\end{equation}
\newpage
\section{Domain Partitions of Geometric Spaces}
In graphic form, the predictive value equations are displayed as follows, with the positive predictive curve in blue, and the negative predictive curve in red:
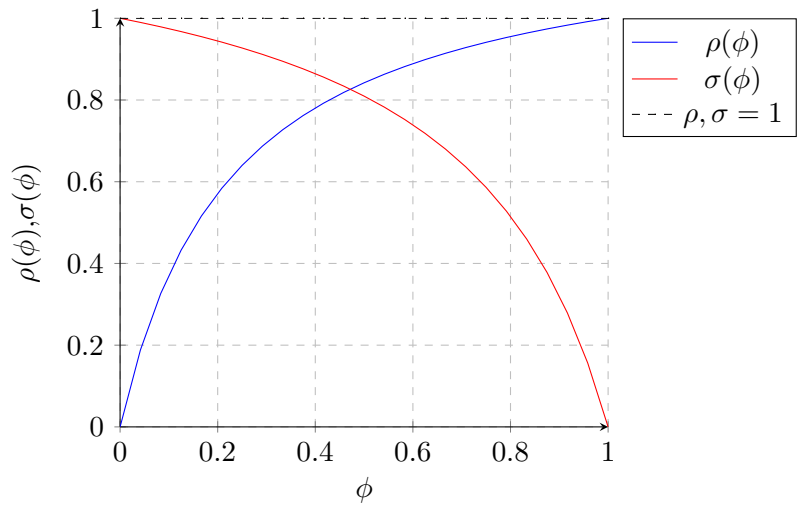
\begin{center}
\begin{tikzpicture}
	\begin{axis}[
    axis lines = left,
    xlabel = $\phi$,
	ylabel = {$\rho(\phi)$,$\sigma(\phi)$},    
     ymin=0, ymax=1,
    legend pos = outer north east,
     ymajorgrids=true,
     xmajorgrids=true,
    grid style=dashed,
    width=8cm,
    height=7cm,
     ]
	\addplot [
	domain= 0:1,
	color= blue,
	]
	{(0.80*x)/((0.80*x+(1-0.85)*(1-x))};

\addplot [	
	domain= 0:1,
	color= red,
	]
	{0.85*(1-x)/((1-0.8)*x+0.85*(1-x))};
	
\addplot+ [
dashed,
domain= 0:1,	
color = black,
mark size = 0pt
 ]
	{1};
	\addlegendentry{$\rho(\phi)$}
	\addlegendentry{$\sigma(\phi)$}
	\addlegendentry{$\rho,\sigma=1$}
	\end{axis}
\end{tikzpicture}
\end{center}

As depicted above, both predictive value curves are continuous, and have domains [0-1] and ranges [0-1]. These curves intersect at a single point within the domain, which is defined as a function of parameters in (4) and (5). In effect, this intersection point, $\phi_i$, partitions the geometric space $\pi(\phi)$ of the predictive curves into two areas: a negative-dominant partition (NDP) and a positive-dominant one (PDP). The negative-dominant partition splits the domain of the geometric space to such values where $\sigma(\phi)$ $>$ $\rho(\phi)$, and vice-versa such that:
\begin{align*}
\pi(\phi)=
\begin{cases} 
      NDP & \text{if } \sigma(\phi) > \rho(\phi), {} dom = [0,\phi_i( \\
       0 & \text{if } \sigma(\phi) = \rho(\phi), {} dom = {} [\phi_i] \\
      PDP & \text{if } \sigma(\phi) < \rho(\phi), {} dom = {} )\phi_i,1] 
\end{cases}
\end{align*}
\\
\
Analytically, we determine the area of partition by obtaining the difference in the definite integral of each curve to and from the intersection point $\phi_i$, such that:

\begin{large}
\begin{equation}
NDP = \int_{0}^{\phi_i} \sigma(\phi) d\phi - \int_{0}^{\phi_i} \rho(\phi) d\phi 
\end{equation}
\end{large}
\begin{large}
\begin{equation}
PDP = \int_{\phi_i}^{1} \rho(\phi) d\phi - \int_{\phi_i}^{1} \sigma(\phi) d\phi 
\end{equation}
\end{large}
\newpage
\subsection{Intersectionality $\phi_i$ of the Predictive Curves}
We can determine the prevalence/pre-test probability level $\phi_i$ where the predictive curves intersect by equating formulas (4) and (5) and isolating $\phi$:

\begin{large}
\begin{equation}
\frac{a\phi}{a\phi+(1-b)(1-\phi)} = \frac{b(1-\phi)}{\phi(1-a)+b(1-\phi)}
\end{equation}
\end{large}
Breaking up the fractions:
\begin{large}
\begin{equation}
{a\phi}[{\phi(1-a)+b(1-\phi)}]=[{a\phi+(1-b)(1-\phi)}][b(1-\phi)]
\end{equation}
\end{large}
Expanding the parentheses:
\begin{large}
\begin{equation}
{a\phi}[{(\phi-a\phi)+(b-b\phi)}]=[a\phi+(1-\phi-b+b\phi)](b-b\phi)
\end{equation}
\end{large}
Re-arranging the terms:
\begin{large}
\begin{equation}
{a\phi}{(\phi-a\phi)+{a\phi}(b-b\phi)}=a\phi(b-b\phi)+(b-b\phi)(1-\phi-b+b\phi)
\end{equation}
\end{large}
Removing ${a\phi}(b-b\phi)$ on either side:
\begin{large}
\begin{equation}
{a\phi}(\phi-a\phi)=(b-b\phi)(1-\phi-b+b\phi)
\end{equation}
\end{large}
We can expand the equations to obtain:
\begin{large}
\begin{equation}
{a\phi^2-a^2\phi^2}=b-b\phi-b^2+b^2\phi-b\phi+b\phi^2+b^2\phi-b^2\phi^2
\end{equation}
\end{large}
Factoring out $\phi$:
\begin{large}
\begin{equation}
(a-b-a^2+b^2)\phi^2 + (2b-2b^2)\phi - (b-b^2)= 0
\end{equation}
\end{large}

We thus obtain a simple quadratic equation of the form $ax^2+bx+c=0$, the solution of which is given by the following relationship: $\frac{-b\pm \sqrt{b^2-4ac}}{2a}$, provided $a^2-b^2-a+b\ne \:0$, which would imply a = b. Given that the domain of the predictive value function is [0,1], we take the negative value of the rooted term to determine the intersection that falls within the domain.

\begin{large}
\begin{equation}
\phi_i=\frac{-b^2+b-\sqrt{ab\left(ab-a+1-b\right)}}{a^2-b^2-a+b}
\end{equation}
\end{large}

Should a tests' sensitivity and specificity be equal, we can overcome nullity in the denominator of (25) by increasing the specificity $b$ by $db$ to obtain an approximate estimation of the intercept point.

\begin{large}
\begin{equation}
b \Rightarrow b +db 
\end{equation}
\end{large}

\subsection{Graphic Representation of Geometric Partitions}
Graphically, we depict the geometric partitions of the screening curves as follows, in an example case where a = 0.80 and b = 0.95, and where the NDP is represented in red and the PDP in blue:

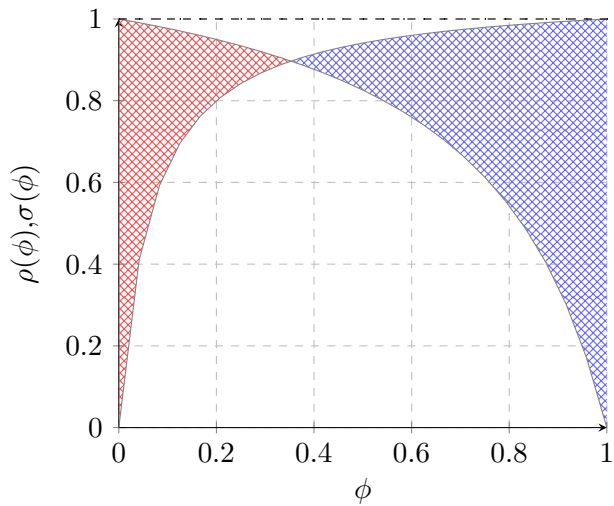
\begin{center}
\begin{tikzpicture}
	\begin{axis}[
    axis lines = left,
    xlabel = $\phi$,
	ylabel = {$\rho(\phi)$,$\sigma(\phi)$},    
     ymin=0, ymax=1,
    legend pos = outer north east,
     ymajorgrids=true,
     xmajorgrids=true,
    grid style=dashed,
    width=8cm,
    height=7cm,
     ]
	\addplot [
	domain= 0:1,
	color= gray,
	name path = A
	]
	{(0.80*x)/((0.80*x+(1-0.95)*(1-x))};

\addplot [	
	domain= 0:1,
	color= gray,
	name path = B
	]
	{0.95*(1-x)/((1-0.8)*x+0.95*(1-x))};

\addplot+ [
dashed,
domain= 0:1,	
color = black,
mark size = 0pt
 ]
	{1};
	
	\addplot fill between[ 
    of = A and B, pattern=crosshatch,
    split, 
    every even segment/.style = {pattern color=red!70, pattern=crosshatch},
    every odd segment/.style  = {pattern color=blue!60, pattern=crosshatch}
  ];

	\end{axis}
\end{tikzpicture}
\end{center}

We can thus see that equation (12) applies well when we seek to enhance the positive predictive value in cases where the pretest probabilities lie in the NDP area. What if in contrast we wanted to enhance the negative predictive value of a test for a condition whose pre-test probability lies beyond $\phi_i$, in the PDP? Using Bayes' theorem, we need to now modify the $\sigma(\phi)$ equation in (5). Once again, following the same derivation as depicted in Theorem 1, this time for $P(\neg T|\neg D)$, we obtain:

\begin{large}
\begin{equation}
\sigma(\phi)= \frac{(1-\phi)\displaystyle\prod_{i=1}^{n}b_n}{(1-\phi)\displaystyle\prod_{i=1}^{n}b_n+\phi\displaystyle\prod_{i=1}^{n}(1-a_n)}
\end{equation}
\end{large}

With this in mind, what happens if during a finite sequence R of negative test ($t$) iterations, be it from a single test or different tests, one stumbles upon a positive test prior to attaining the desired negative predictive value?

\begin{center}
R = $\{t_\pm\}_{t=0}^{n_i} = \lbrace t_{1-}, t_{2-}, t_{3+},...n_i \rbrace$
\end{center}

Akin to all the derivations previously discussed in this work, the ensuing positive predictive value is described by the following equation:

\begin{large}
\begin{equation}
\lambda(\phi)= \frac{(1-a_{1-})(1-a_{2-})...a_{n+}\phi}{(1-a_{1-})(1-a_{2-})...a_{n+}\phi+b_{1-}b_{2-}...(1-b_{n+})(1-\phi)} 
\end{equation}
\end{large}

where $a_{x-}$ and $b_{x-}$ are the sensitivity and specificity of the $xth$ negative test and $a_{x+}$ and $b_{x+}$ are the like parameters of the positive test.
Simplified, the equation becomes:

\begin{large}
\begin{equation}
\lambda(\phi)= \frac{\phi a_{n+}\displaystyle\prod_{i=a_{1-}}^{a_{(n-1)-}}(1-a_{n-})}{\phi a_{n+}\displaystyle\prod_{i=a_{1-}}^{a_{(n-1)-}}(1-a_{n-})+[(1-\phi)(1-b_{n+})]\displaystyle\prod_{i=b_{1-}}^{b_{(n-1)-}}b_{n-}}
\end{equation}
\end{large}

\section{Conclusion}
The equations derived in this work provide deductive, quantitative proofs regarding the impact of sequential and orthogonal testing on the predictive value of screening tests in different clinical scenarios where screening is undertaken. Their clinical utility is best observed in conditions with low pre-test probability where single tests are insufficient to achieve clinically significant predictive values and likewise, in clinical scenarios with a high pre-test probability where confirmation of disease status is critical. Application of the methods herein described in clinical situations is needed to provide empirical proof of their validity and usefulness.
\newpage

\bibliographystyle{plain}
\bibliography{references}

\end{document}